\documentclass[aps,prl,floatfix,twocolumn,footinbib,superscriptaddress]{revtex4-1}
\usepackage{epsfig}
\usepackage{float}
\usepackage{epstopdf} 
\usepackage{braket}
\usepackage[utf8]{inputenc}
\usepackage{amsmath}
\usepackage{amssymb}
\usepackage{esint}
\usepackage{xcolor}
\usepackage{bbm}

\begin{document}

\global\long\def\avg#1{\langle#1\rangle}

\global\long\def\p{\prime}

\global\long\def\dg{\dagger}

\global\long\def\ket#1{|#1\rangle}

\global\long\def\bra#1{\langle#1|}

\global\long\def\proj#1#2{|#1\rangle\langle#2|}

\global\long\def\inner#1#2{\langle#1|#2\rangle}

\global\long\def\tr{\mathrm{tr}}

\global\long\def\pd#1#2{\frac{\partial#1}{\partial#2}}

\global\long\def\spd#1#2{\frac{\partial^{2}#1}{\partial#2^{2}}}

\global\long\def\der#1#2{\frac{d#1}{d#2}}

\global\long\def\im{\imath}

\global\long\def\S{\mathcal{S}}

\global\long\def\A{\mathcal{A}}

\global\long\def\F{\mathcal{F}}

\global\long\def\E{\mathcal{E}}

\global\long\def\As{{^{\sharp}}\hspace{-1mm}\mathcal{A}}

\global\long\def\Fs{{^{\sharp}}\hspace{-0.7mm}\mathcal{F}}

\global\long\def\Es{{^{\sharp}}\hspace{-0.5mm}\mathcal{E}}

\global\long\def\EsG{{^{\sharp}}\hspace{-0.5mm}\mathcal{E}_{G}}

\global\long\def\EsB{{^{\sharp}}\hspace{-0.5mm}\mathcal{E}_{B}}

\global\long\def\FsG{{^{\sharp}}\hspace{-0.5mm}\F_{G}}

\global\long\def\FsB{{^{\sharp}}\hspace{-0.5mm}\F_{B}}

\global\long\def\Fd{{^{\sharp}}\hspace{-0.7mm}\mathcal{F}_{\delta}}

\global\long\def\EG{\mathcal{E}_{G}}

\global\long\def\EB{\mathcal{E}_{B}}

\global\long\def\O{\mathcal{O}}

\global\long\def\SgF{\S d\F}

\global\long\def\SgEF{\S d\left(\E/\F\right)}

\global\long\def\U{\mathcal{U}}

\global\long\def\V{\mathcal{V}}

\global\long\def\H{\mathbf{H}}

\global\long\def\SO{\Pi_{\S}}

\global\long\def\PO{\hat{\Pi}_{\S}}

\global\long\def\SSH{\tilde{\Pi}_{\S}}

\global\long\def\EO{\Upsilon_{k}}

\global\long\def\ESH{\Omega_{k}}

\global\long\def\HSF{\mathbf{H}_{\S\F}}

\global\long\def\HSEF{\mathbf{H}_{\S\E/\F}}

\global\long\def\HS{\mathbf{H}_{\S}}

\global\long\def\ES{H_{\S}(t)}

\global\long\def\ESo{H_{\S}(0)}

\global\long\def\EgF{H_{\SgF} (t)}

\global\long\def\EgE{H_{\S d\E}(t)}

\global\long\def\EgEF{H_{\SgEF} (t)}

\global\long\def\EF{H_{\F}(t)}

\global\long\def\EFo{H_{\F}(0)}

\global\long\def\ESF{H_{\S\F}(t)}

\global\long\def\ESEF{H_{\S\E/\F}(t)}

\global\long\def\ESSEF{H_{\tilde{\S}\S\E/\F}(t)}

\global\long\def\EEFo{H_{\E/\F}(0)}

\global\long\def\EEF{H_{\E/\F}(t)}

\global\long\def\HPB{H(\PB)}

\global\long\def\MI{I\left(\S:\F\right)}

\global\long\def\aMI{\left\langle \MI\right\rangle _{\Fs}}

\global\long\def\BS{\Pi_{\S} }

\global\long\def\PB{\hat{\Pi}_{\S} }

\global\long\def\QD{\mathcal{D}\left(\Pi_{\S}:\F\right)}
\global\long\def\QD{\mathcal{D}(\Pi_{\S}:\F)}

\global\long\def\QDp{\mathcal{D}\left(\PB:\F\right)}
\global\long\def\QDpIL{\mathcal{D}(\PB:\F)}

\global\long\def\JI{J\left(\Pi_{\S}:\F\right)}

\global\long\def\CI{H\left(\F\left|\Pi_{\S}\right.\right)}

\global\long\def\CIp{H\left(\F\left|\PB\right.\right)}

\global\long\def\CS{\rho_{\F\left|s\right.}}

\global\long\def\CSu{\tilde{\rho}_{\F\left|s\right.}}

\global\long\def\CSp{\rho_{\F\left|\hat{s}\right.}}

\global\long\def\CEF{H_{\F\left|s\right.}}

\global\long\def\CEFp{H_{\F\left|\hat{s}\right.}}

\global\long\def\psiz{\ket{\psi_{\E\left|0\right.\hspace{-0.4mm}}}}

\global\long\def\psio{\ket{\psi_{\E\left|1\right.\hspace{-0.4mm}}}}

\global\long\def\psiinner{\inner{\psi_{\E\left|0\right.\hspace{-0.4mm}}}{\psi_{\E\left|1\right.\hspace{-0.4mm}}}}

\global\long\def\QDz{\boldsymbol{\delta}\left(\S:\F\right)_{\left\{  \sigma_{\S}^{z}\right\}  }}

\global\long\def\NQD{\bar{\boldsymbol{\delta}}\left(\S:\F\right)_{\BS}}

\global\long\def\EFS{H_{\F\left| \BS\right. }(t)}

\global\long\def\EFSM{H_{\F\left| \left\{  \ket m\right\}  \right. }(t)}

\global\long\def\Hol{\chi\left(\Pi_{\S}:\F\right)}
\global\long\def\HolIL{\chi (\Pi_{\S}:\F) }

\global\long\def\Holp{\chi\left(\PB:\F\right)}
\global\long\def\HolpIL{\chi ( \PB:\F )}

\global\long\def\ch{\raisebox{0.5ex}{\mbox{\ensuremath{\chi}}}_{\mathrm{Pointer}}}

\global\long\def\rhoS{\rho_{\S}(t)}

\global\long\def\rhoSo{\rho_{\S}(0)}

\global\long\def\rhoSF{\rho_{\S\F} (t)}

\global\long\def\rhoSgEF{\rho_{\SgEF} (t)}

\global\long\def\rhoSgF{\rho_{\SgF} (t)}

\global\long\def\rhoF{\rho_{\F}(t)}

\global\long\def\rhoFp{\rho_{\F}(\pi/2)}

\global\long\def\LE{\Lambda_{\E}(t)}

\global\long\def\LEc{\Lambda_{\E}^{\star}(t)}

\global\long\def\LEij{\Lambda_{\E}^{ij}(t)}

\global\long\def\LF{\Lambda_{\F}(t)}

\global\long\def\LFij{\Lambda_{\F}^{ij} (t)}

\global\long\def\LFc{\Lambda_{\F}^{\star}(t)}

\global\long\def\LEF{\Lambda_{\E/\F} (t)}

\global\long\def\LEFij{\Lambda_{\E/\F}^{ij}(t)}

\global\long\def\LEFc{\Lambda_{\E/\F}^{\star}(t)}

\global\long\def\Lkij{\Lambda_{k}^{ij}(t)}

\global\long\def\Hb{H}

\global\long\def\kE{\kappa_{\E}(t)}

\global\long\def\kEF{\kappa_{\E/\F}(t)}

\global\long\def\kF{\kappa_{\F}(t)}

\global\long\def\ts{t=\pi/2}

\global\long\def\QCB{\bar{\xi}_{QCB}}

\global\long\def\mc#1{\mathcal{#1}}

\global\long\def\MD{\lambda}

\global\long\def\up{\mathord{\uparrow}}

\global\long\def\down{\mathord{\downarrow}}

\global\long\def\Cku{\rho_{k\left|\up\right.}}

\global\long\def\Ckd{\rho_{k\left|\down\right.}}

\global\long\def\f{\mathcal{J}}

\global\long\def\onlinecite#1{\cite{#1}}

\newcommand{\todo}[1]{\textcolor{red}{#1}}
\newcommand{\RN}[1]{\uppercase\expandafter{\romannumeral#1}}

\title{Revealing the emergence of classicality in nitrogen-vacancy centers}


\author{T. K. Unden}
\affiliation{Institute for Quantum Optics, Ulm University, Albert-Einstein-Allee 11, Ulm 89081, Germany}
\author{D. Louzon}
\affiliation{Institute for Quantum Optics, Ulm University, Albert-Einstein-Allee 11, Ulm 89081, Germany}
\affiliation{Racah Institute of Physics, The Hebrew University of Jerusalem, Jerusalem 91904, Israel}

\author{M. Zwolak}
\affiliation{Biophysics Group, Microsystems and Nanotechnology Division, Physical Measurement Laboratory, National Institute of Standards and Technology, Gaithersburg,
Maryland 20899, U.S.A.}

\author{W. H. Zurek}
\affiliation{Theory Division, Los Alamos National Laboratory, Los Alamos, NM, 87545, U.S.A.}

\author{F. Jelezko}
\affiliation{Institute for Quantum Optics, Ulm University, Albert-Einstein-Allee 11, Ulm 89081, Germany}
\affiliation{Center for Integrated Quantum Science and Technology (IQ$^\text{{st}}$), Ulm University, 89081 Germany}

\begin{abstract}
{
The origin of classical reality in our quantum world is a long-standing mystery. Here, we examine a nitrogen vacancy center in diamond evolving in the presence of its magnetic nuclear spin environment which is formed by the natural appearance of carbon $^{13}C$ atoms in the diamond lattice, to study quantum Darwinism -- the proliferation of information about preferred quantum states throughout the world via the environment. This redundantly imprinted information accounts for the perception of objective reality, as it is independently accessible by many without perturbing the system of interest. To observe this process, we implement a novel dynamical decoupling scheme that enables the measurement/control of several nuclear spins (the environment $\E$) interacting with a nitrogen vacancy (the system $\S$). 
Our experiment demonstrates that, in course of decoherence of $\S$, redundant information is indeed imprinted onto $\E$, giving rise to incipient classical objectivity -- a consensus recorded in redundant copies, and available from the fragments of the nuclear spin environment $\E$, about the state of $\S$. This provides the first laboratory verification of the process responsible for the emergence of the objective classical world from the underlying quantum substrate.

}
\end{abstract}

\maketitle

Quantum Darwinism -- a theoretical framework for describing the emergence of the classical world from the quantum -- recognizes that the environment is a communication channel through which observers acquire information. This upgrades the role of the environment from the one it had in decoherence theory (i.e., just suppressing quantum superpositions) and provides a framework for understanding and quantifying the emergence of the objective classical world~\cite{Ollivier04-1,Ollivier05-1,Blume-Kohout2005,zur09,Zwolak13-1,Zurek14-1,Zwolak14,Pawel15,zwolak16,Zwolak17-1,Gerardo18}. In the process of decohering a system, the environment selectively acquires information about certain system states -- the pointer states~\cite{Zurek81-1} that are resistant to decoherence -- and transmits it to observers who can then find out about $\S$ independently and indirectly via $\E$. In our world the same photon environment that contributes to decoherence simultaneously and inherently gives rise to our perception of objective states of fundamentally quantum systems. These are the pointer states that survive the interaction with the environment and promulgate information about themselves into the world.

\begin{figure}
\centerline{\psfig{file=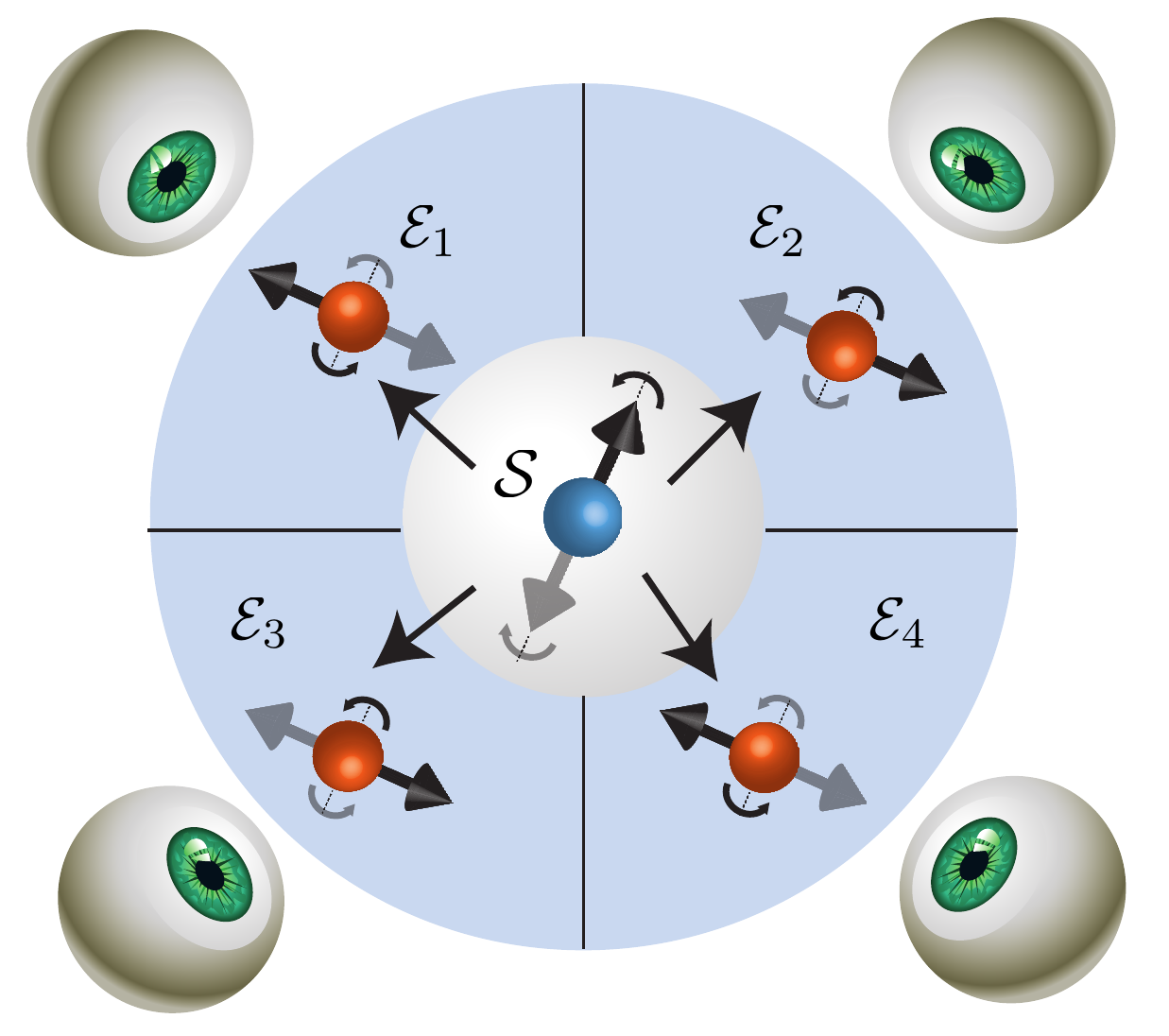, width=0.8\linewidth}}
\caption{The nuclear spin environment as a quantum communication channel. A central electronic spin -- the system $\S$ -- is surrounded by multiple nuclear spins $\E_k$ comprising the environment $\E$. The environment spins are effectively isolated from each other due to their weak spin-spin interaction. The interaction between the central spin and an individual nuclear spin is mediated by the hyperfine interaction and depends on the relative position of each spin. The hyperfine interaction strength is therefore different for each nuclear spin. The environment decoheres the system and, in the process, each of its components is rotated into a new state (black and grey arrows) conditional on the central spin state. Multiple observers (eyes) can access different environment spins and thus independently deduce the state of the system.}
\label{fig:1}
\end{figure}

This process of selective proliferation of information responsible for the emergence of the classical world is most effective on the macroscopic level (when ``order'' Avogadro's number of environment components interact with the system), but it has to be studied in the microscopic quantum domain. Decohering interactions of a class that includes the photon environment, as well as spin and other models (so called ``pure decoherence''), universally give rise to redundant imprinting of information which in turn gives rise to objective classical reality~\cite{Zwolak14,zwolak16}.  Central spin systems, in particular, offer ideal test cases to observe this emergence in action and even control it, see Fig.~\ref{fig:1}. Such experiments are still rather challenging. Real systems are inhomogeneous, which means measurement and control requires addressing disparate components. Moreover, central system and the subsystems of the model environmnetn tend to interact with everything and, together with spectral broadening, this makes identifying and selecting the most relevant interactions difficult. Nitrogen-vacancy (NV) centers provide an interesting setup where some of these issues can be solved, as we will show.  

\begin{figure*}
\centerline{\psfig{file=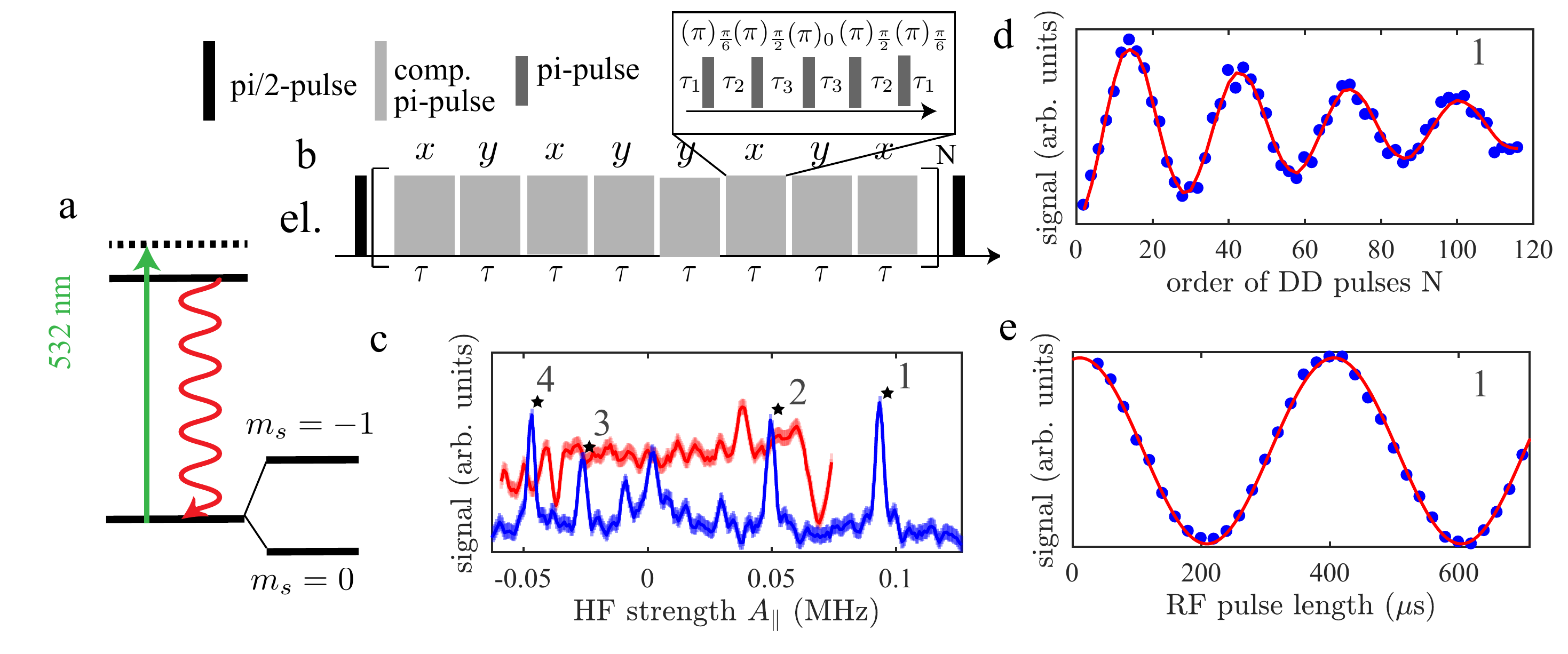, width=1\linewidth}}
\caption{Experimental control of the nuclear-electronic spin system. (a) Energy level diagram of the electronic spin of a NV center. The orbital configuration can be optically excited with green laser light and the passage through the excited state is manifested by red fluorescence. Each orbital state carries a spin triplet ($S=1$) manifold. Spin-dependent non-radiative decay can be used for optical spin detection and efficient spin initialization of the ground state spin sublevel $\ket{m_s = 0}$. In this work we focus on the two-level-system specified by the ground state spin sublevels $\ket{m_s=0} \equiv \ket{\up}$ and $\ket{m_s=-1} \equiv \ket{\down}$. (b) The adaptive XY8$^N$  sequence. The DD sequence is a train of composite $\pi$-pulses with a single pulse duration of $\tau$ and an alternating orthogonal phase (here $x$ and $y$) for robustness. Each composite $\pi$-pulse is a symmetric sequence of five microwave $\pi$-pulses (inset) with different pulse phases to achieve robustness against single pulse imperfections. (c) Measured spectrum (blue) when the interpulse spacing $\tau = 1/(\omega_L+A_{\parallel}/2)$ of an AXY$^{16}$ sequence varies, where $\omega_L$ specifies the bare Larmor frequency of nuclear carbon spins determined by the external, applied magnetic field (here, $\approx 440$ G). For comparison, the result of a standard XY8$^{16}$ spectrum is shown in red. The solid lines are smoothed data and the light blue/red shaded regions represent one standard deviation. The four strongest coupled nuclear spins are marked by stars with corresponding numbers and the parallel hyperfine coupling strengths $93.5$ kHz, $49.5$ kHz, $-26.3$ kHz and $-47.1$ kHz are identified. (d) NV spin mediated Rabi oscillations of a single nuclear spin. The interpulse spacing $\tau$ is tuned to the Larmor period of carbon spin 1 and the order $N$ of the AXY sequence is increased. The red curve is the result of a simulation, when the measured hyperfine values (see section \RN{3} in the SI) are taken into account. (e) Rabi oscillation of nuclear spin 1 driven by a RF field. AXY sequences are used for initialization and readout of the nuclear spin (see section \RN{1} in the SI for more information). The solid curve is a cosine fit corresponding to a sum of squares error (SSE) of $2.4\cdot10^{-4}$. Errors are smaller than the data points in (d,e).}
\label{fig:2}
\end{figure*}

We will focus on the single electron spin in an NV center~\cite{DOHERTY20131,PSSA:PSSA200671403}, Fig.~\ref{fig:2}a, embedded in a room-temperature diamond environment that carries nuclear $^{13}$C spins (with the natural abundance of 1.1 \%). The diamond sample is grown via Chamical Vapor Deposition (CVD) and single NV center are introduced into the diamond from residual nitrogen in the CVD plasma. Such a platform - a central electron spin coupled to nuclear ancilla spins - has also been studied in previous experiments in different contexts \cite{Taminiau14,Hirose16,Zaiser16}. In the secular approximation~\cite{Childress281}, the Hamiltonian is 
\begin{equation}
\H = 2\pi S_z\sum_k A^k_\parallel I^k_z ,
\label{eq:H0}
\end{equation}
where $S_z = \proj{\up}{\up}$ is a shifted $z$-operator for the electron spin, $I^k_z$ is the spin-1/2 operator for the nuclear spin $k$ and $A^k_\parallel$ the parallel component of the hyperfine interaction (HF) vector $\vec{A}^k$. This is of the pure decoherence form, where environment components interact with the system and do not interact with each other~\cite{Zwolak14,zwolak16,Riedel12-1}. The eigenstates of $S_z$, are the so-called pointer states of the system~\cite{Zurek81-1}. The states that are not perturbed by the environment even though their superpositions decohere. For an initial state where the electron spin is in a non-classical quantum superposition, $\ket{+}=\ket{\up}+\ket{\down}$, and in a product state with the environment spins (individually in an initialized state $\ket{\phi_k}$), 
\begin{equation}
\ket{\psi(0)} = \ket{+}_\S \otimes \left[ \bigotimes_k \ket{\phi_k} \right],
\label{eq:Psi0}
\end{equation}
the state after evolving for a time $t$ is
\begin{equation}
\ket{\psi(t)} = \ket{\up}_\S \otimes \left[ \bigotimes_k \ket{\phi_{k|\up}} \right] + \ket{\down}_S \otimes \left[ \bigotimes_k \ket{\phi_{k|\down}} \right].
\label{eq:GHZ}
\end{equation}
The superposition in the system has ``branched out'' into the environment, creating correlations with the nuclear spins via conditional rotations into the states $\ket{\phi_{k|\hat{s}}}$ with $\hat{s}=\up,\down$ the pointer states of the system (the $m_s =0$ and $-1$ states of the NV center, respectively, see Fig.~\ref{fig:2}a). 

When $\ket{\phi_k} = \ket{+}$, $A^k_\parallel=A_\parallel$, and $t=1/(2A_\parallel)$, the state in Eq.~\eqref{eq:GHZ} is a GHZ state, where each environment spin holds a perfect record of the pointer state, i.e., the conditional states $\ket{\phi_{k|\up}}$ and $\ket{\phi_{k|\down}}$ are orthogonal and thus the system's pointer state can be inferred exactly. Under more general conditions, the state is GHZ-like and each spin only holds a partial record of the system's state. In either case, the information can be quantified by the quantum mutual information between the system $\S$ and a fragment $\F$ of the environment, 
\begin{equation} \label{eq:MI}
\MI = \ES + \EF - \ESF ,
\end{equation}
where $H_\A=-\tr \rho_\A \log_2 \rho_A$ is the von Neumann entropy of subsystem $\A$. This decomposes into classical and quantum components~\cite{Zwolak13-1}, 
\begin{equation} \label{eq:QC}
\MI = \Hol + \QD .
\end{equation}
The first component is the Holevo quantity \cite{Holevo73,Nielsen11}
\begin{equation}
\Hol = \EF - \sum_{s} p_s \CEF(t) ,
\end{equation}
which upper bounds the classical information communicated by a quantum channel, i.e., here, information about the observable $\Pi_{\S}$ on the system $\S$ communicated by an environment fragment $\F$. The second component, $\QD$, gives the quantum discord~\cite{Zurek00-1,Ollivier02-1,Henderson01-1}. The quantity $\CEF$ is the entropy of $\F$ conditioned on outcome $s$ in $\S$ (with probability $p_s$). 

In principle, one can examine the information about any observable of $\S$, but under decoherence it is information about the pointer states of $\S$, $\PO$ ($S_z$ in our case), that is imprinted on $\F$~\cite{Ollivier04-1,Zwolak13-1}. In what follows, we will determine $\HolpIL$ in a natural setting. 
We focus on the Holevo information because its complement in the equation for mutual information   -- quantum discord $\QDpIL$ -- describes correlations between $S$ and $F$ that cannot be shared by observers \cite{Zurek13}, and, hence, cannot help establish objective reality.
We thus focus only on $\HolpIL$ and will discuss possibilities for obtaining $\QDpIL$ afterward.

In the case of the generation of a perfect GHZ state (see artificial, experimental creation in section \RN{4} of the SI), the Holevo information is 1 bit for any fragment of the environment: The original system's state is perfectly decohered and each environment spin carries a record of the system's pointer state. Thus, several observers which each intercept one spin from the environment, they can all independently determine the pointer state of the system. This is the notion of redundancy, that there are (in this ideal case) $\Es$ copies of the information about the system in the environment of size $\Es$. Departing from ideality, the redundancy, $R_\delta$, will be $\Es/\Fd$ where $\Fd$ is the size of the typical fragment required to obtain
\begin{equation} \label{eq:Red}
\avg{\HolpIL} \ge (1-\delta) \HPB .
\end{equation}
That is, the fragment size, on average, to get more than $\HPB$ of the missing information about $\S$. The quantity $\delta$ is the information deficit -- the finite precision one has to pay for lack of ideality. 


It is clear that to observe this process in the laboratory, one either has to perform full quantum state tomography or, to see that there is redundant information, address the individual nuclear spins. State-of-the-art technology uses Dynamical Decoupling (DD) to tackle issues such as these. However, selectivity in a spectrally dense environment is still a difficult task. Here, we implement a novel DD protocol, theoretically proposed in Refs.~\cite{Cas2015,Cas17}, to both identify the spin environment and to control individual parts of it. Like well-established DD sequences such as CPMG~\cite{MAUDSLEY1986488} or XY8~\cite{gull90}, the protocol employs repetitive central spin flips via a microwave (MW) drive, where the inter pulse spacing determines the frequency of the control window. However, the new protocol, the adaptive XY8 (AXY8) sequence, establishes a robust control of individual nuclear spins mediated by the central electron spin by arbitrarily shaping the DD control-filter. This refocuses undesired noise, allowing for the identification and control of individual nuclear spins. 

More specifically, control of the filter design is supplied by replacing each single spin flip by a train of five pulses, see the inset of Fig.~\ref{fig:2}b. An alternating rotation axis (phase) of the MW pulses permits a robust operation in the presence of pulse errors. In addition, time evolution during the pulse train models an arbitrary filter response, where the evolution times $\tau_1$, $\tau_2$ and $\tau_3$ are numerically calculated with a specific filter function (see section \RN{1} in the SI). In case that the nuclear Zeeman energy is much larger than the hyperfine coupling strength, the process is modeled by the effective Hamiltonian~\cite{Cas2015}
\begin{equation}
H^k = \frac{1}{2}f_{DD}A_\perp \left(S_z -\frac{1}{2}\right) I^k_x,
\label{eq:EM1}
\end{equation}
when the interpulse spacing $\tau$ matches the corresponding Larmor frequency of nuclear spin $k$. $I_x^k$ is the corresponding nuclear spin-$1/2$ operator in $x$-direction and $f_{DD}$ a variable parameter, determining the DD control filter. 

The effective interaction strength $\frac{f_{DD}\A_\perp}{2}$ is mediated by the perpendicular HF coupling $A_\perp$, which is here determined by the magnetic dipole-dipole interaction. Instead of a constant interaction strength, determined by $A_\perp$, a weaker effective strength can be modeled without the necessity of using higher harmonics~\cite{Childress281,Tam14}, which are more vulnerable to pulse error, to achieve individual nuclear spin addressing. Further experiments (see section \RN{1} of the SI) perfomed with different filter coefficients confirm the concept of the AXY sequence and artefacts coming from contributions of higher harmonics can be neglegted due to a large detuning.  An AXY spectrum (with $f_{DD}=0.2$) of the nuclear spin environment is shown in Fig.~\ref{fig:2}c (blue curve). Due to electron-nuclear spin entanglement governed by Eq.~\eqref{eq:EM1}, four stronger coupled nuclear spins and additional, more weakly coupled ones can be identified by their different parallel HF interaction strength, when the interpulse spacing $\tau$ varies. The effective coupling strength was here reduced by about a factor of five compared to the dipolar HF interaction strength determined by the register geometry. The result of the typical, non-adaptive XY8 sequence (red curve) doesn't show the features. The HF interaction strength is too strong, and therefore the resonances too broad to identify individual spins. The rising and falling of quantum correlations between the electronic spin and a nuclear spin is shown in Fig.~\ref{fig:2}d, when the repetition $N$ of the pulse sequence is increased, while the pulse spacing is kept constant and on resonance with the Larmor period of nuclear spin one. In addition, the result in Fig.~\ref{fig:2}e shows a Rabi oscillation of nuclear spin one induced by a resonant radio frequency (RF) field, when a iSwap gate \cite{Cas17} based on the AXY sequence is used for nuclear spin initialization and readout (see more information in section \RN{1} of the SI). 

\begin{figure*}
\centerline{\psfig{file=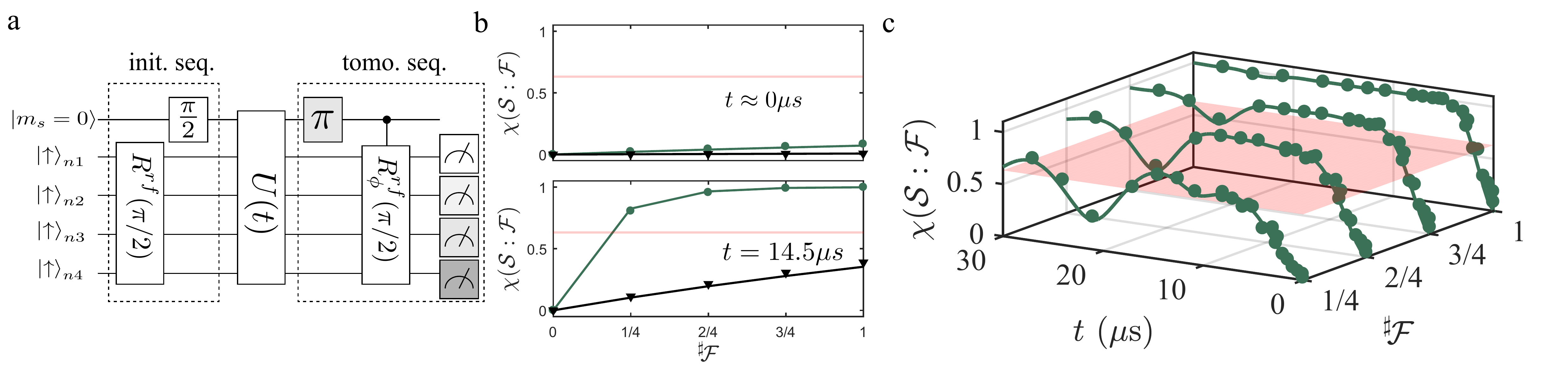, width=1\linewidth}}
\caption{The emergence of redundancy for an NV center being naturally decohered by its environment. (a) The NV spin is first initialized optically and its polarization swapped to each individual nuclear spin by a repetitive process (not shown). Two $\frac{\pi}{2}$-pulses transform the product state into a product state of $\ket{+}$ states. These then evolve under the direct HF interaction between the NV center and nuclear spins ($U(t)$). Single nuclear spin tomography in the electronic subspace $\ket{m_s=-1}$ is performed by a selective $\frac{\pi}{2}$-pulse mediated by a weak, resonant RF pulse ($R^{rf}_{\phi}$). In addition, multiple measurements are performed with different RF pulse phases $\phi$ to determine the phase of the nuclear spin superposition. An optional $\pi$-pulse in front of the last RF pulse can be applied for nuclear spin tomography in the electronic $\ket{m_s=0}$ subspace. The state of a single nuclear spin is in the end swapped to the NV spin and an optical readout follows. (b) Holevo information versus fraction size for two different free evolution times. (c) Holevo information, $\HolpIL$, versus the environment fragment size $\Fs$ and free evolution time $t$. The solid curves in (b) and (c) show the results of simulations with and without imperfect initial polarization. The dynamics in the simulation are governed by the Hamiltonian $H_0$, Eq.~\eqref{eq:H0}. The semi-transparent red lines in (b) and the plane in (c) indicate an information deficit of $1/e$, i.e., $I = (1-1/e) H_\S$. Errors are smaller than the data points.}
\label{fig:4}
\end{figure*}

In previous work, redundancy was created artificially via the construction of a GHZ state, which was achieved for example recently with photonic simulators~\cite{Ciampini18-1,Chen18-1}. In addition, work in the field of quantum non-demolition (QND)  measurements \cite{Nogues99,Gleyzes07,Lupascu07,Neumann542} were also able to create highly redundant states by consecutive two-body scattering. However, in our everyday world, redundancy appears as a consequence of natural interactions between an $\S$ and $\E$ initially out of equilibrium. To observe this in NV centers, we allow in the following the system to evolve freely in the presence of the natural HF interaction. Because the effect of the nuclear spin environment on the central electron spin is dominated by the nuclear spins close to the central spin, we concentrate on the four strongest coupled nuclear spins (see Fig.~\ref{fig:2}c). The experimental protocol is shown in Fig.~\ref{fig:4}a. We first initialize the system into the out-of-equilibrium product state, Eq.~\eqref{eq:Psi0}, with $\ket{\phi_k}=\ket{+}$. This is followed by the free evolution of $\S\E$ of duration $t$ according to the HF Hamiltonian, Eq.~\eqref{eq:H0}. To determine the classical correlations of fragments, $\F$, of the environment with $\S$, tomography is applied by an electron spin-selective nuclear RF pulse with variable phase $\phi$, which rotates only an individual nuclear spin (see the section \RN{2} and \RN{4} in the SI for more information). Nuclear spin initialization and readout is achieved by a nuclear spin selective iSwap gate mediated by the AXY sequence.

Figure~\ref{fig:4}b shows the Holevo information versus fraction size for two different evolution times and Fig.~\ref{fig:4}c shows the full data set. When errors happen during the nuclear spin polarization as well as tomography, and both sorts of errors are corrected in the analysis (see section \RN{2} of the SI), the results are shown in dark green. When only the errors of an imperfect tomography are corrected in the analysis, the data is presented in black. Here we focus on the dark green data set. At short times, there is essentially no information in fragments or even the whole environment, as initially the $\S\E$ state is a product state. As time develops, however, information is rapidly transferred into the environment. At a time of 14.5 $\mu s$ even a single nuclear spin captures nearly complete information, to within an information deficit of $1/e$ bits. In other words, all the four most strongly coupled nuclear spins have nearly a complete record of the system's pointer state. This redundancy is reflected in the presence of a plateau in the Holevo information versus fragment size. We note also that the timescale of information rising (on the order of several $\mu$s) is in good agreement with the NV spin coherence time measured by a Ramsey experiment, see section \RN{7} in the SI. With increasing time, though, small fragments will see information flow back into the system. This is due to the fact that for individual spins the conditional states first get rotated away from each other and then back towards each other. How fast this conditional rotation of an individual nuclear spin will take place is mediated by the HF coupling. For example the observed decrease in the Holevo information at a time of about 20 $\mu$s is due the conditional rotations of nuclear spin one, two and four which fulfill almost $2\pi$-cycles. For sufficiently large number of spins with random interaction strengths, even this information flow into a single spin will be one way on average. For larger fragment sizes, the information tends more and more to be one way, although there will still be periods of recurrence for long enough waiting times. When the experimental data are not normalized with respect to the initial degree of polarization (black data), redundancy is suppressed due to the lower -- but nonzero except for cases of measure zero~\cite{Zwolak14,zwolak16} -- information capacity of the ``hazy'' environmental fragments~\cite{Zwolak09,Zwolak10-1}. 

We have exhibited the emergence of redundancy under decoherence using NV centers. Our study also provides insights to the reason for the selective banishing of quantum information: the initially superposition becomes encoded into the inaccessible global quantum correlations. This global coherence leaves a signature in the quantum mutual information ($\QDpIL$, the counterpart to $\HolpIL$ in Eq.~\eqref{eq:QC}) in the form of an ``uptick'' a sharp upward turn of the mutual information on the plateau when the fragment size near the total environment size~\cite{Zwolak13-1}. 
This can be observed artificially e.g., in isolated photonic simulators~\cite{Ciampini18-1,Chen18-1} and similar settings where couplings between $\S$ and elements of $\E$ are artificially controlled. Within natural settings, though, interactions with inaccessible environment components, as well as imperfect readout/initialization of the accessible environment components, make observing the uptick very challenging. Indeed, the inaccessibility of the uptick due to the interactions with many environment components is what makes our everyday world classical~\cite{Zwolak13-1}. Thus, it is no surprise that it is difficult to measure in naturally decohering systems. Further refinement of the DD technique and samples, together with low temperature measurements, may make this uptick accessible. This will motivate future experiments in NV centers embedded in moderately $^{13}$C enriched diamond (see section \RN{8} in the SI) to observe large amounts of redundancy.

Our results provide the first laboratory demonstration of quantum Darwinism in action in a natural environment. This demonstration required implementing a novel DD protocol. The process by which nuclear spin decoherence of NV centers gives rise to incipient classical objectivity is analogous to the one that occurs when photons scatter from objects in our macroscopic world. In both cases, flagrantly non-classical (e.g., non-local) quantum superpositions are embedded in larger environment, initially out of equilibrium. Interactions with the environment select certain preferred (pointer) states of the system, decohering their superpositions and proliferating accessible information about such einselected states into the world, thus relegating non-redundant quantum correlations to inaccessible regions of the Hilbert space. Our work shows that already on the atomic scale there is evidence of the process that -- in everyday settings, and for much larger environments -- leads to the emergence of classicality. The appearance of objective, classical states accessible to indirect measurements is anticipated by processes that take place already in small environments, and it simply gets more difficult to avoid classicality as the environment size grows. This straightforward and purely quantum account of the origins of the classical in our quantum Universe suggests other approaches to the quantum-to-classical transition (gravitational collapse, etc.) are not necessary to describe the emergence of our objective, classical world.\\

\section*{\label{sec:Acknowledgement} Acknowledgement}
We thank Jorge Casanova, Zhenyu Wang, Liam P. McGuinness, and C. Jess Riedel for helpful discussions. This work was supported by the DoE LDRD program at Los Alamos National Laboratory, FQX, ERC, VW Stiftung, BW Stiftung, DFG and BMBF.  WHZ acknowledges partial support by the Foundational Questions Institute grant FQXi-1821, and Franklin Fetzer Fund, a donor advised fund of the Silicon Valley Community Foundation. 

\nocite{bib,Cas2015,Souza4748,Cas17,Laraoui15,Baum85,Gaerttner17,Zwolak09,Zwolak10-1,Audenaert07,Zwolak14,zwolak16,Nizovtsev18}

\bibliographystyle{apsrev}
\bibliography{reference}

\end{document}


\title{Revealing the emergence of classicality in nitrogen-vacancy centers}


\global\long\def\avg#1{\langle#1\rangle}

\global\long\def\p{\prime}

\global\long\def\dg{\dagger}

\global\long\def\ket#1{|#1\rangle}

\global\long\def\bra#1{\langle#1|}

\global\long\def\proj#1#2{|#1\rangle\langle#2|}

\global\long\def\inner#1#2{\langle#1|#2\rangle}

\global\long\def\tr{\mathrm{tr}}

\global\long\def\pd#1#2{\frac{\partial#1}{\partial#2}}

\global\long\def\spd#1#2{\frac{\partial^{2}#1}{\partial#2^{2}}}

\global\long\def\der#1#2{\frac{d#1}{d#2}}

\global\long\def\im{\imath}

\global\long\def\S{\mathcal{S}}

\global\long\def\A{\mathcal{A}}

\global\long\def\F{\mathcal{F}}

\global\long\def\E{\mathcal{E}}

\global\long\def\As{{^{\sharp}}\hspace{-1mm}\mathcal{A}}

\global\long\def\Fs{{^{\sharp}}\hspace{-0.7mm}\mathcal{F}}

\global\long\def\Es{{^{\sharp}}\hspace{-0.5mm}\mathcal{E}}

\global\long\def\EsG{{^{\sharp}}\hspace{-0.5mm}\mathcal{E}_{G}}

\global\long\def\EsB{{^{\sharp}}\hspace{-0.5mm}\mathcal{E}_{B}}

\global\long\def\FsG{{^{\sharp}}\hspace{-0.5mm}\F_{G}}

\global\long\def\FsB{{^{\sharp}}\hspace{-0.5mm}\F_{B}}

\global\long\def\Fd{{^{\sharp}}\hspace{-0.7mm}\mathcal{F}_{\delta}}

\global\long\def\EG{\mathcal{E}_{G}}

\global\long\def\EB{\mathcal{E}_{B}}

\global\long\def\O{\mathcal{O}}

\global\long\def\SgF{\S d\F}

\global\long\def\SgEF{\S d\left(\E/\F\right)}

\global\long\def\U{\mathcal{U}}

\global\long\def\V{\mathcal{V}}

\global\long\def\H{\mathbf{H}}

\global\long\def\SO{\Pi_{\S}}

\global\long\def\PO{\hat{\Pi}_{\S}}

\global\long\def\SSH{\tilde{\Pi}_{\S}}

\global\long\def\EO{\Upsilon_{k}}

\global\long\def\ESH{\Omega_{k}}

\global\long\def\HSF{\mathbf{H}_{\S\F}}

\global\long\def\HSEF{\mathbf{H}_{\S\E/\F}}

\global\long\def\HS{\mathbf{H}_{\S}}

\global\long\def\ES{H_{\S}(t)}

\global\long\def\ESo{H_{\S}(0)}

\global\long\def\EgF{H_{\SgF} (t)}

\global\long\def\EgE{H_{\S d\E}(t)}

\global\long\def\EgEF{H_{\SgEF} (t)}

\global\long\def\EF{H_{\F}(t)}

\global\long\def\EFo{H_{\F}(0)}

\global\long\def\ESF{H_{\S\F}(t)}

\global\long\def\ESEF{H_{\S\E/\F}(t)}

\global\long\def\ESSEF{H_{\tilde{\S}\S\E/\F}(t)}

\global\long\def\EEFo{H_{\E/\F}(0)}

\global\long\def\EEF{H_{\E/\F}(t)}

\global\long\def\HPB{H(\PB)}

\global\long\def\MI{I\left(\S:\F\right)}

\global\long\def\aMI{\left\langle \MI\right\rangle _{\Fs}}

\global\long\def\BS{\Pi_{\S} }

\global\long\def\PB{\hat{\Pi}_{\S} }

\global\long\def\QD{\mathcal{D}\left(\Pi_{\S}:\F\right)}
\global\long\def\QD{\mathcal{D}(\Pi_{\S}:\F)}

\global\long\def\QDp{\mathcal{D}\left(\PB:\F\right)}
\global\long\def\QDpIL{\mathcal{D}(\PB:\F)}

\global\long\def\JI{J\left(\Pi_{\S}:\F\right)}

\global\long\def\CI{H\left(\F\left|\Pi_{\S}\right.\right)}

\global\long\def\CIp{H\left(\F\left|\PB\right.\right)}

\global\long\def\CS{\rho_{\F\left|s\right.}}

\global\long\def\CSu{\tilde{\rho}_{\F\left|s\right.}}

\global\long\def\CSp{\rho_{\F\left|\hat{s}\right.}}

\global\long\def\CEF{H_{\F\left|s\right.}}

\global\long\def\CEFp{H_{\F\left|\hat{s}\right.}}

\global\long\def\psiz{\ket{\psi_{\E\left|0\right.\hspace{-0.4mm}}}}

\global\long\def\psio{\ket{\psi_{\E\left|1\right.\hspace{-0.4mm}}}}

\global\long\def\psiinner{\inner{\psi_{\E\left|0\right.\hspace{-0.4mm}}}{\psi_{\E\left|1\right.\hspace{-0.4mm}}}}

\global\long\def\QDz{\boldsymbol{\delta}\left(\S:\F\right)_{\left\{  \sigma_{\S}^{z}\right\}  }}

\global\long\def\NQD{\bar{\boldsymbol{\delta}}\left(\S:\F\right)_{\BS}}

\global\long\def\EFS{H_{\F\left| \BS\right. }(t)}

\global\long\def\EFSM{H_{\F\left| \left\{  \ket m\right\}  \right. }(t)}

\global\long\def\Hol{\chi\left(\Pi_{\S}:\F\right)}
\global\long\def\HolIL{\chi(\Pi_{\S}:\F)}

\global\long\def\Holp{\chi\left(\PB:\F\right)}
\global\long\def\HolpIL{\chi(\PB:\F)}

\global\long\def\ch{\raisebox{0.5ex}{\mbox{\ensuremath{\chi}}}_{\mathrm{Pointer}}}

\global\long\def\rhoS{\rho_{\S}(t)}

\global\long\def\rhoSo{\rho_{\S}(0)}

\global\long\def\rhoSF{\rho_{\S\F} (t)}

\global\long\def\rhoSgEF{\rho_{\SgEF} (t)}

\global\long\def\rhoSgF{\rho_{\SgF} (t)}

\global\long\def\rhoF{\rho_{\F}(t)}

\global\long\def\rhoFp{\rho_{\F}(\pi/2)}

\global\long\def\LE{\Lambda_{\E}(t)}

\global\long\def\LEc{\Lambda_{\E}^{\star}(t)}

\global\long\def\LEij{\Lambda_{\E}^{ij}(t)}

\global\long\def\LF{\Lambda_{\F}(t)}

\global\long\def\LFij{\Lambda_{\F}^{ij} (t)}

\global\long\def\LFc{\Lambda_{\F}^{\star}(t)}

\global\long\def\LEF{\Lambda_{\E/\F} (t)}

\global\long\def\LEFij{\Lambda_{\E/\F}^{ij}(t)}

\global\long\def\LEFc{\Lambda_{\E/\F}^{\star}(t)}

\global\long\def\Lkij{\Lambda_{k}^{ij}(t)}

\global\long\def\Hb{H}

\global\long\def\kE{\kappa_{\E}(t)}

\global\long\def\kEF{\kappa_{\E/\F}(t)}

\global\long\def\kF{\kappa_{\F}(t)}

\global\long\def\ts{t=\pi/2}

\global\long\def\QCB{\bar{\xi}_{QCB}}

\global\long\def\mc#1{\mathcal{#1}}

\global\long\def\MD{\lambda}

\global\long\def\up{\mathord{\uparrow}}

\global\long\def\down{\mathord{\downarrow}}

\global\long\def\Cku{\rho_{k\left|\up\right.}}

\global\long\def\Ckd{\rho_{k\left|\down\right.}}

\global\long\def\f{\mathcal{J}}

\global\long\def\onlinecite#1{\cite{#1}}

\newcommand{\todo}[1]{\textcolor{red}{#1}}

\title{Supplementary Information -- Revealing the emergence of classicality in nitrogen-vacancy centers}

\author{T. Unden}
\affiliation{Institute for Quantum Optics, Ulm University, Albert-Einstein-Allee 11, Ulm 89081, Germany}

\author{D. Louzon}
\affiliation{Institute for Quantum Optics, Ulm University, Albert-Einstein-Allee 11, Ulm 89081, Germany}
\affiliation{Racah Institute of Physics, The Hebrew University of Jerusalem, Jerusalem 91904, Israel}

\author{M. Zwolak}
\affiliation{Center for Nanoscale Science and Technology, National Institute of Standards and Technology, Gaithersburg,
Maryland 20899, U.S.A.}

\author{W. H. Zurek}
\affiliation{Theory Division, Los Alamos National Laboratory, Los Alamos, NM, 87545, U.S.A.}

\author{F. Jelezko}
\affiliation{Institute for Quantum Optics, Ulm University, Albert-Einstein-Allee 11, Ulm 89081, Germany}
\affiliation{Center for Integrated Quantum Science and Technology (IQ$^\text{{st}}$), Ulm University, 89081 Germany}

\maketitle

\renewcommand\thefigure{S\arabic{figure}}

\section{Experimental Setup and Control}

The diamond sample (IIa) is grown via Chemical Vapor Deposition (CVD). The isotopic composition is $98.9$ \% $^{12}C$. The concentration of carbon isotope nuclear spins ($1.1$ \% $^{13}$C) gives a significant probability to find multiple weakly coupled $^{13}$C in the frozen core of a NV center. Single NV center are introduced into the diamond from residual nitrogen in the CVD plasma.

For optical manipulation and detection of single NV centers in bulk diamond we use a home-built confocal microscope and for initialization and readout of the NV ground state spin green laser excitation ($532$ nm). An avalanche photodiode detects the corresponding fluorescence and a permanent magnet gives the static field ($\approx440$ G). We drive the electronic (NV) and nuclear ($^{13}$C) spin transitions via MW and RF fields via a thin copper wire close to the NV center and an arbitrary-wave form generator. The MW/RF phase gives the rotation axis of a pulse. We insert a $\frac{\pi}{2}$ phase-shift to get rotations around the $y$-axis. For manipulation, the typical Rabi frequencies are $40$ MHz for the electron spin rotations and $5$ kHz for nuclear spin rotations.

The DD sequence proposed in Ref.~\cite{Cas2015} provides control. Adjusting the timing $(\tau_1,\tau_2, \tau_3)$ and therefore the interpulse spacing $\tau$ can give an arbitrary filter-function for the DD sequence. We focus on the first harmonic and set the other contributions up to $4^{th}$ order to zero. The strength of the filter-function is set by the filter coefficients  $f_1=f_{DD},f_2=0,f_3=0,f_4=0$ and we solve the equations (see Ref.~\cite{Cas2015})
\begin{equation}
f_k = \frac{4}{\pi k} \left[\sum_{j=1}^{2} (-1)^j\left[(-1)^k-1\right]\sin\left(2\pi k \theta_j\right)+\sin\left(k\frac{\pi}{2}\right)\right]
\end{equation}
for $k=1-4$ numerically to get the corresponding pair of times 
\begin{align}
\tau_1 &= \frac{\theta_1}{\omega_{DD}}\\
\tau_2 &= \frac{\theta_2-\theta_1}{\omega_{DD}}\\  
\tau_3 &= \frac{1}{4\omega_{DD}}.
\end{align}
The frequency $\omega_{DD}=\frac{1}{2\tau}$ is the center of the control filter. Any effective coupling 
\begin{equation}
\frac{f_{DD}A_\perp}{2} \in \frac{1}{\pi}A_\perp\left(-8\cos(\frac{\pi}{9})+4,8\cos(\frac{\pi}{9})-4\right)
\end{equation}
is possible, where $A_\perp$ is the perpendicular HF coupling strength of an individual nuclear spin. 

\begin{figure}[H]
\centerline{\psfig{file=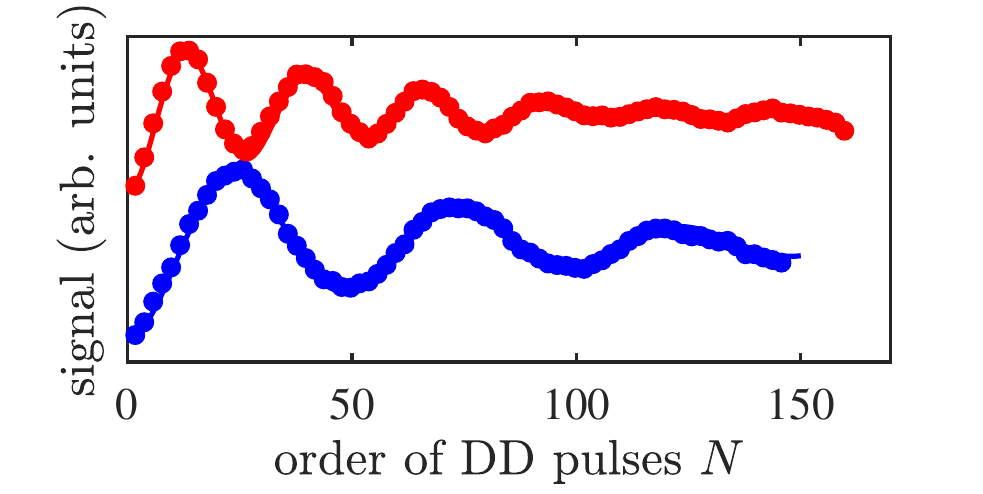, width=0.5\linewidth}}
\caption{Electron spin coherence versus the DD pulse number when the pulse spacing is adjusted to the resonance of carbon spin 1. The entangling operation is via the AXY pulse sequence with filter coefficient $f_{DD}=0.1$ (blue) and doubled strength $f_{DD}=0.2$ (red). Solid lines are the result of a simulation, assuming dephasing and the measured hyperfine coupling strength (see Table 1). Errors are smaller than the data points.}
\label{fig:S1}
\end{figure}

\begin{figure}[H]
\centerline{\psfig{file=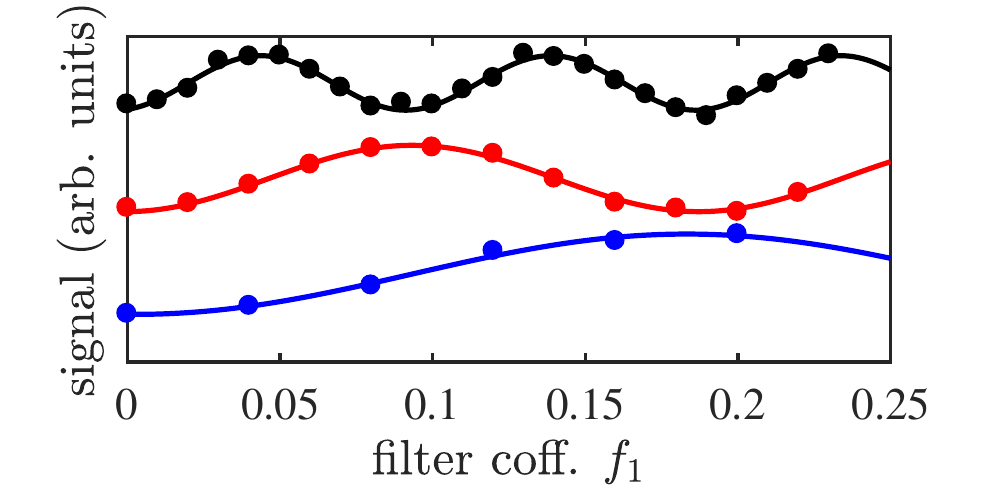, width=0.5\linewidth}}
\caption{Electron spin coherence versus different filter strength. The entangling operation is via AXY-8 (blue)/16 (red)/32 (black). The solid lines are cosine fits (SSE from up to down of $4\cdot10^{-4}$, $6\cdot10^{-4}$ and $29\cdot10^{-4}$ ). Errors are smaller than the data points.}
\label{fig:S2}
\end{figure}

In Fig.~\ref{fig:S1}, we show the coherent interaction between the electron and nuclear spin $\E_1$ for two different strengths of the filter-function. When the effective coupling increases from $\frac{0.1A_\perp}{2}$ to $\frac{0.2A_\perp}{2}$, the frequency of the oscillation doubles. Figure~\ref{fig:S2} shows the coherent interaction between the electron and nuclear spin $\E_1$ for a fixed AXY repetition number and increasing effective coupling for the cases $N=8,16,32$. A filter strength of about $\frac{0.2\cdot A_\perp}{2}$ was used. For a certain $N$, we fine-tune $f_{DD}$ to achieve two qubit $\pi/2$-rotations.

For further control, nuclear spin initialization and individual measurement via optical readout of the NV electronic spin is necessary. We achieve both by using an entangling gate mediated by $H^k$ (see main text), in combination with an asymmetric version~\cite{Souza4748} of it, to implement an iSwap gate ($e^{i\frac{\pi}{2}\sigma_xI_x}e^{i\frac{\pi}{2}\sigma_yI_y}$)~\cite{Cas17}. For the strongest coupled nuclear spin, it takes about $50\mu s$ to perform a swap operation, and for the weakest coupled one about $80\mu s$. We implement the entangling gate $R=e^{iH^k\frac{\pi}{2}}$ for each nuclear spin by carefully adjusting the repetition $N$ of the AXY sequence and the interpulse spacing $\tau$. To observe for example the nuclear Rabi oscillations shown in the main text, first the electron spin was optically polarized, then polarization was swapped to nuclear spin 1 with additional reinitialization of electron spin, application of the RF pulse follows and nuclear spin information is optically read out after a second iSwap gate. A typical gate fidelity of an iSwap gate is on the order of $0.8$ and is mainly limited by the electron coherence time of about 0.7 ms. Specific HF coupling values of the four nuclear spin system can be found in section \ref{sec.hf}. 

\section{Analysis}

The Holevo information requires the diagonal (in the $\PB$ basis of the system) component of the $\S\E$ density matrix. We thus can take the density matrix to be of the form $\rho_{\S\E}= p_{\up} \proj{\up}{\up} \otimes \rho_{\E\left|\up\right.} + p_{\down} \proj{\down}{\down} \otimes \rho_{\E\left|\down\right.}$, where the off-diagonal components in $\PB$ are not present (e.g., $\proj{\up}{\down}$ and $\proj{\down}{\up}$) and $\rho_{\E\left|\hat{s}\right.}=\bigotimes_k \rho_{\E_k\left|\hat{s}\right.}$ are the conditional nuclear density matrices in the corresponding electronic spin (pointer) sublevel $\hat{s}$. We determine the parameters of the nuclear density matrices (see for example section \ref{sec.GHZ}) by fitting the experimental data with simulation (and $p_{\up} = p_{\down} = 1/2$). The Lindbladian for the simulation consists of the coherent HF interaction and a pure dephasing term associated with the HF coupling to additional, weakly coupled nuclear spins. We precisely measure in advance the parallel and perpendicular components of the HF coupling to the four core spins and the dephasing rate. By performing this fitting, we correct the $\S\E$ density matrix for errors occurring during the tomography process. To correct also the impact of imperfect nuclear spin initialization, the individual nuclear spin density matrices are normalized, to take on the form 
\begin{equation}
\rho_{\E_k} = \begin{pmatrix}
\rho_{11} & \rho_{12}/P_k\\
\rho_{21}/P_k & \rho_{22}
\end{pmatrix} ,
\end{equation}
where $P_k$ is the initial degree of nuclear spin polarization (after the swap operations). This normalizes the experimental data to the optical contrast of NV Rabi oscillations. 

\section{Hyperfine coupling strength}\label{sec.hf}

Table~\ref{tab:HF} shows the hyperfine coupling strength of the different nuclei. The parallel component is estimated by analysis of AXY spectra and the perpendicular component by analyzing oscillations (see Fig.~\ref{fig:S1}) caused by resonant interaction when the interaction time is increased. 

\begin{table}[H]
\caption{Hyperfine interaction strength of the register.}
\begin{center}
\begin{tabular}{ c c c }
  \hline
  \textbf{nuc. spin} & \textbf{$\frac{A_\parallel}{2\pi}$ (kHz)} & \textbf{$\left| \frac{A_\perp}{2\pi}\right|$ (kHz)} \\ 
  \hline
  1 & $\phantom{-}93.5$ & $45.8$ \\
  2 & $\phantom{-}49.5$ & $35.3$ \\
  3 & $-26.3$ &$22.0$\\
  4 & $-47.1$ & $42.5$\\
  \hline
\end{tabular}
\end{center}
\label{tab:HF}
\end{table}

\section{Creation of an electronic-nuclear GHZ state}\label{sec.GHZ}

\begin{figure}[H]
\centerline{\psfig{file=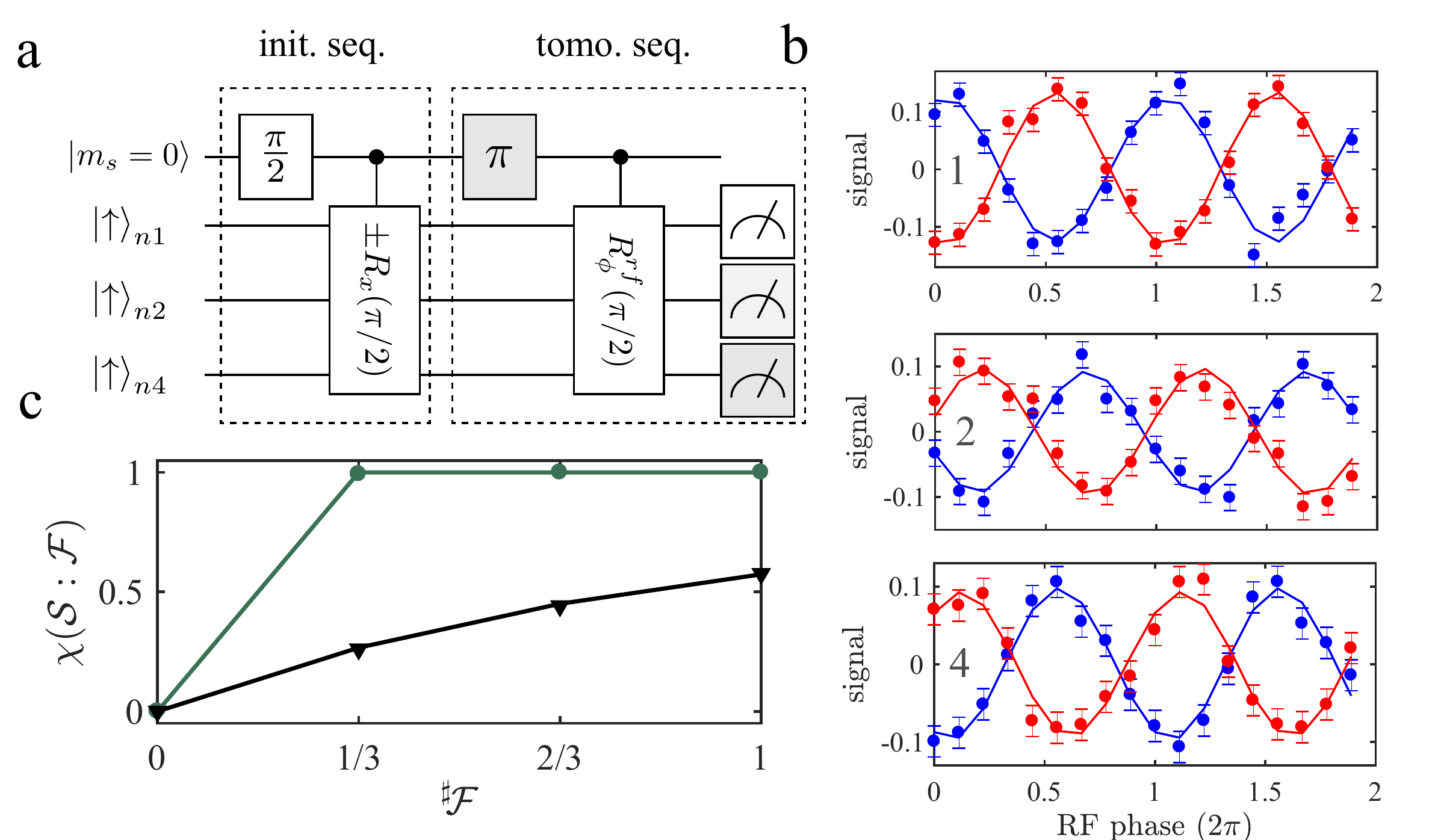, width=0.8\linewidth}}
\caption{Storing the central spin state redundantly in a three nuclear spin environment. (a) Protocol for creation and readout of the redundant state. The NV spin is first initialized and its polarization swapped to each individual nuclear spin by a repetitive process (not shown). An electronic superposition is then created by a $\frac{\pi}{2}$-pulse. Electronic-nuclear spin entanglement is created by a $\frac{\pi}{2}$ nuclear rotation in a direction conditional on the electronic spin state (see main text). Single nuclear spin tomography in the electronic subspace $\ket{m_s=-1}$ is performed by a selective $\frac{\pi}{2}$-pulse mediated by weak, resonant RF pulse. In addition, multiple measurements are performed with a different RF pulse phase $\phi$ to determine the phase of the nuclear spin superposition. An optional $\pi$-pulse in front of the last RF pulses can be applied for nuclear spin tomography in the electronic $\ket{m_s=0}$ subspace. The state of a single nuclear spin is in the end swapped to the NV spin and an optical readout follows. In a single sequence run, a nuclear spin is readout in one of the electronic subspaces. (b) Measurement results corresponding to the sequence shown in (a). The normalized NV fluorescence for each nuclear spin $(1,2,4)$ is presented. Tomography in the $\ket{m_s=0}$ ($\ket{m_s=-1}$) subspace is indicated in blue (red). Solid curves are results of a simulation (see Methods). Error bars represent one standard deviation of the measured data points. (c) Holevo information $\HolpIL$ based on the analysis of the data in (b) when different environment fraction sizes $\Fs$ are taken. The black data is corrected for error happening during tomography step (see the Analysis section) and the dark green data is also corrected for imperfect initial polarization. The solid curves show the result of a simulation considering the entangled state, $\ket{\up}_\S \otimes \ket{+}+\ket{\down}_\S \otimes \ket{-}$, with introduced initial polarization imperfections (black) and without (dark green). Errors are smaller than the data points.} 
\label{fig:3}
\end{figure}

The protocol to create a GHZ state is in Fig.~\ref{fig:3}a. For this demonstration, the three strongest coupled nuclear spins are chosen. We polarize the nuclear spins by first polarizing the electron spin via optical pumping and then swapping the electron polarization to each nuclear spin. This is followed by an electronic $\pi/2$-pulse and entanglement is sequentially created by the application of the entangling gate described in the main text. To observe correlations with a single nuclear spin, we utilize a selective RF $\pi/2$-pulse with variable phase $\phi$ ($R^{rf}_\phi$), which only rotates a single nuclear spin when the NV center is in the $m_s = -1$ state. To have access to the $m_s=0$ state, an optional MW $\pi$-pulse can be applied in front of the RF pulse. The results are shown in Fig.~\ref{fig:3}b for each nuclear spin. By varying the phase of the RF pulse, oscillations occur. For all nuclear spins we observe a $\pi$-phase shift between the data observed in the $m_s = 0$ (in blue) and $m_s = -1$ (in red) state, which indicates the orthogonal preparation of the state . The phase shift between signals of different nuclear spins is due to a different HF tensor with respect to the laboratory frame, defined by the RF-field axis~\cite{Laraoui15}. Here we focus mainly on classical correlations, but the result of an echo kind of sequence shows also the presence of quantum correlations in form of multiple quantum coherences~\cite{Baum85,Gaerttner17}, see next section.  

Figure~\ref{fig:3}c presents the Holevo information versus the fraction of nuclear spins in the fragment. In black we show the analyzed data, when error happening during the tomography sequence are corrected and in dark green the results when also error due to imperfect initial nuclear spin polarization is corrected (details are in the Analysis section). The typical degree of polarization is on the order of $(75\pm 5) \%$ for each nuclear spin and the fidelity for creating the GHZ state is about $40$ \% (see next section). Focusing on the data set shown in green, as soon as just one spin from the environment is intercepted, the Holevo information is already 1 bit. That is, access to a single environment spin already gives the pointer state of the system. Interception of further spins does not increase the Holevo information but only confirms the information from the first. This plateau -- the classical plateau -- signifies the appearance of redundant information. When error due to imperfect initial polarization are not corrected (black data), only the initial rise to the plateau is seen. Due to a non-zero initial entropy, the amount of information a single nuclear spin can store decreases and even when all environmental spins are taken into account, only about half a bit of central spin information can be intercepted. In general this is not true when a larger spin environment is considered~\cite{Zwolak09,Zwolak10-1}. 

\section{Multiple quantum coherence}

\begin{figure}[H]
\centerline{\psfig{file=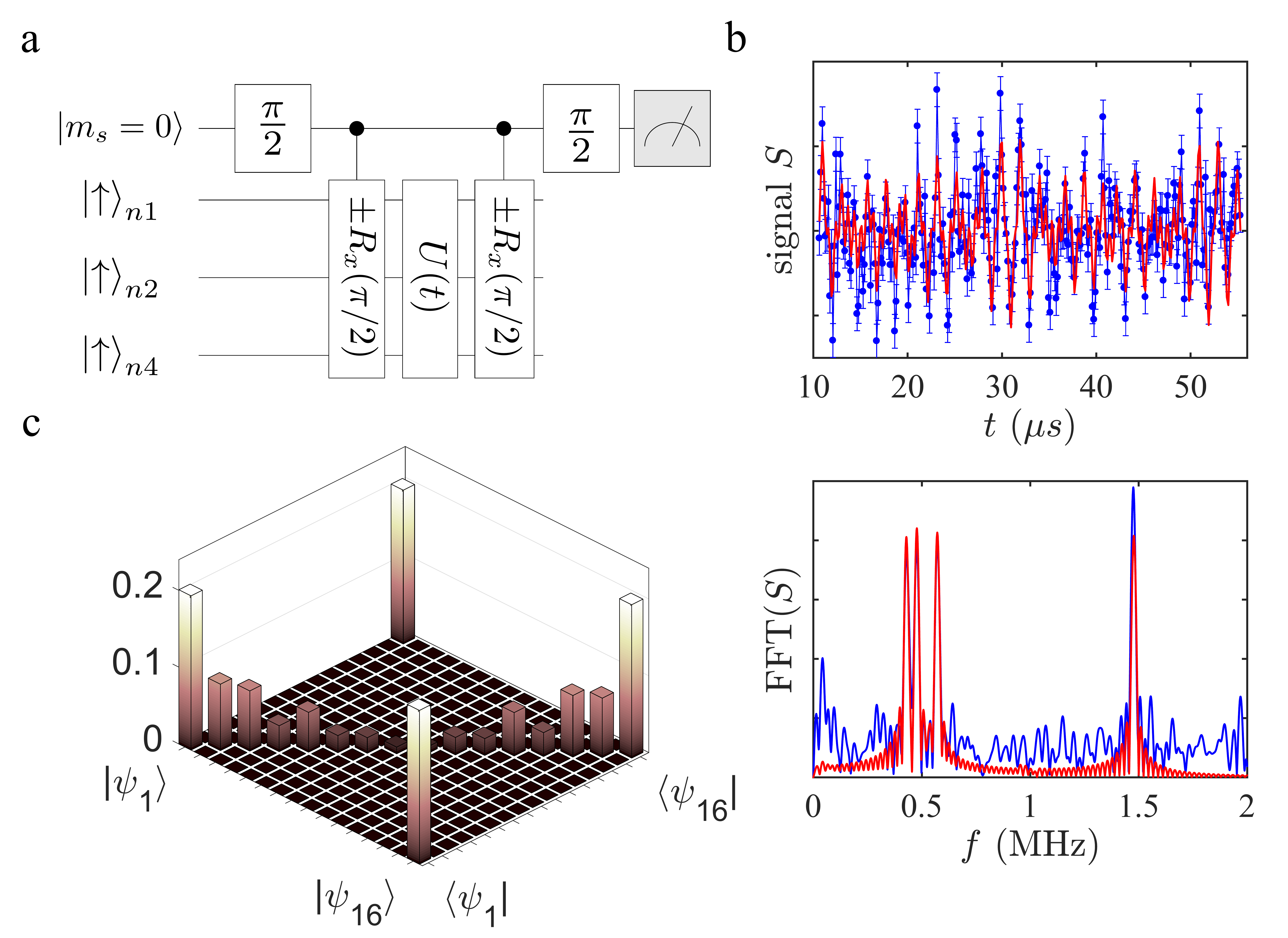, width=0.6\linewidth}}
\caption{Results of the Loschmidt echo. (a) Sequence for testing multiple coherences present in the system. (b) Measurement (blue) and simulation (red) results with corresponding Fast Fourier Transform (FFT). Single Larmor frequency at a magnetic field of about $440G\approx 0.5 MHz$ is visible. The shift of each nuclear spin precession is due hyperfine interaction. Also visible is the sum of individual Larmor frequencies, which indicate multiple coherences present in the system. Error bars represent one standard deviation. (c) Reconstructed density matrix. with $\ket{\psi_1}=\ket{\uparrow}\ket{+}\ket{+}\ket{+}\ket{+}$ and  $\ket{\psi_{16}}=\ket{\downarrow}\ket{-}\ket{-}\ket{-}\ket{-}$.}
\label{fig:S4}
\end{figure}

In Fig.~\ref{fig:S4}a, we show our sequence for testing quantum correlations after creating the redundant GHZ state. It is based on a Loschmidt echo, where the echo is perturbed by a free evolution period in between both entangling gates ($\pm R_x(\frac{\pi}{2})$). By increasing the free evolution period $\tau$, oscillations, Fig.~\ref{fig:S4}b, get visible, which can be attributed to the Larmor precession (about $0.5$ MHz) of the three nuclear spins and the sum of all Larmor frequencies (about $1.5$ MHz), which is a hint for multiple quantum coherences. The free evolution period was composed of a CPMG sequence, where the pulse spacing did not match to nuclear spin transitions. The reconstructed density matrix (shown in Fig.~\ref{fig:S4}c) is the result of a comparison of experimental data with simulated data (that takes only the diagonal and outer off-diagonal entries of the density matrix into account) and yields a state fidelity of $40$ \% with respect to a fully entangled GHZ state. 

\section{Chernoff Information}

\begin{figure}[H]
\centerline{\psfig{file=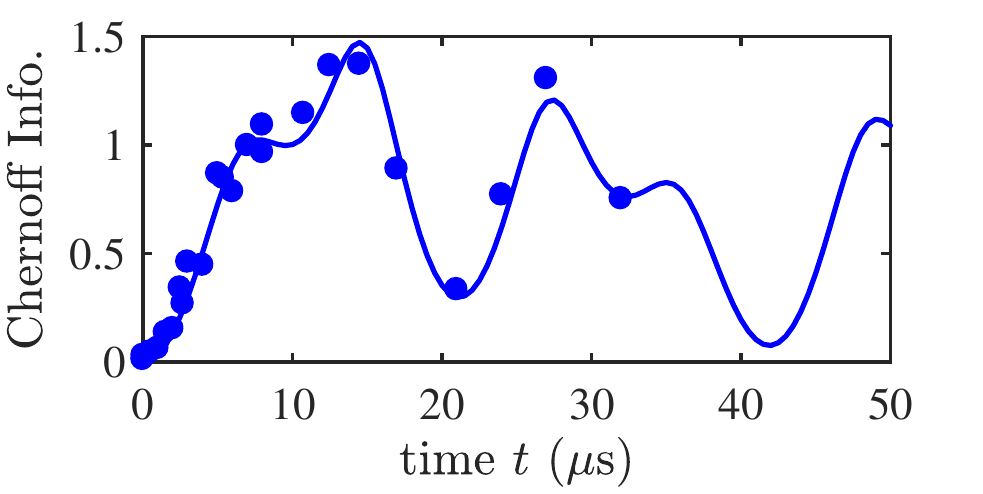, width=0.5\linewidth}}
\caption{Chernoff Information. Solid line corresponds to a simulation based on the measured hyperfine interactions. Errors are smaller than the data points.}
\label{fig:S5}
\end{figure}

In Fig.~\ref{fig:S5}, we show the quantum Chernoff information~\cite{Audenaert07} averaged over the four nuclear spins~\cite{Zwolak14,zwolak16}, i.e., the typical quantum Chernoff information $\QCB$, which sets an upper bound for redundancy rate in the asymptotic (large fragment size) limit. The redundancy is given by $\QCB \Es / \ln 1/\delta$, where $\Es$ is the total number of components in the environment. This equation requires sufficiently precise information (i.e., a small information deficit $\delta$) in order for the asymptotic limit to be accurate, otherwise the redundancy would be incorrectly given a value greater than $\Es$. 

\section{NV Ramsey}

\begin{figure}[H]
\centerline{\psfig{file=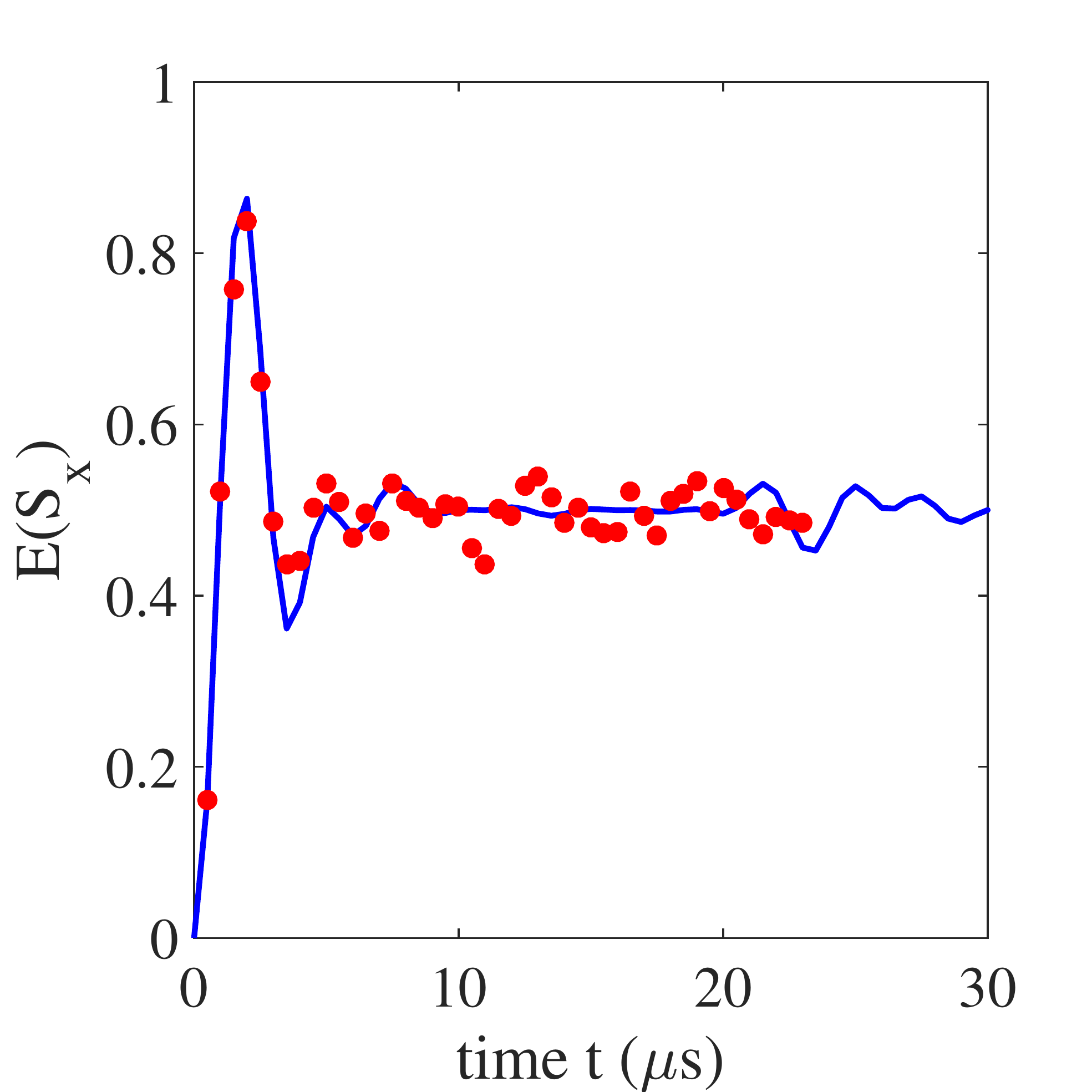, width=0.35\linewidth}}
\caption{NV Ramsay experiment when the nuclear spin register is polarized to the $\ket{+}$ state. Solid line corresponds to a simulation based on the measured hyperfine interactions. Errors are smaller than the data points.}
\label{fig:S3}
\end{figure}

A Ramsay ($\frac{\pi}{2}$-pulse - wait $\tau$ - $\frac{\pi}{2}$-pulse) experiment performed on the electronic spin of NV center is shown in Fig.~\ref{fig:S3}. The four strongest coupled nuclear spins are polarized into the $\ket{+}$-state in this measurement. The initial state is therefore in accordance with the initial state of the experiment shown in the main text. 

\section{Potential redundancy in $^{13}C$ enriched diamonds}

\begin{figure}[H]
\centerline{\psfig{file=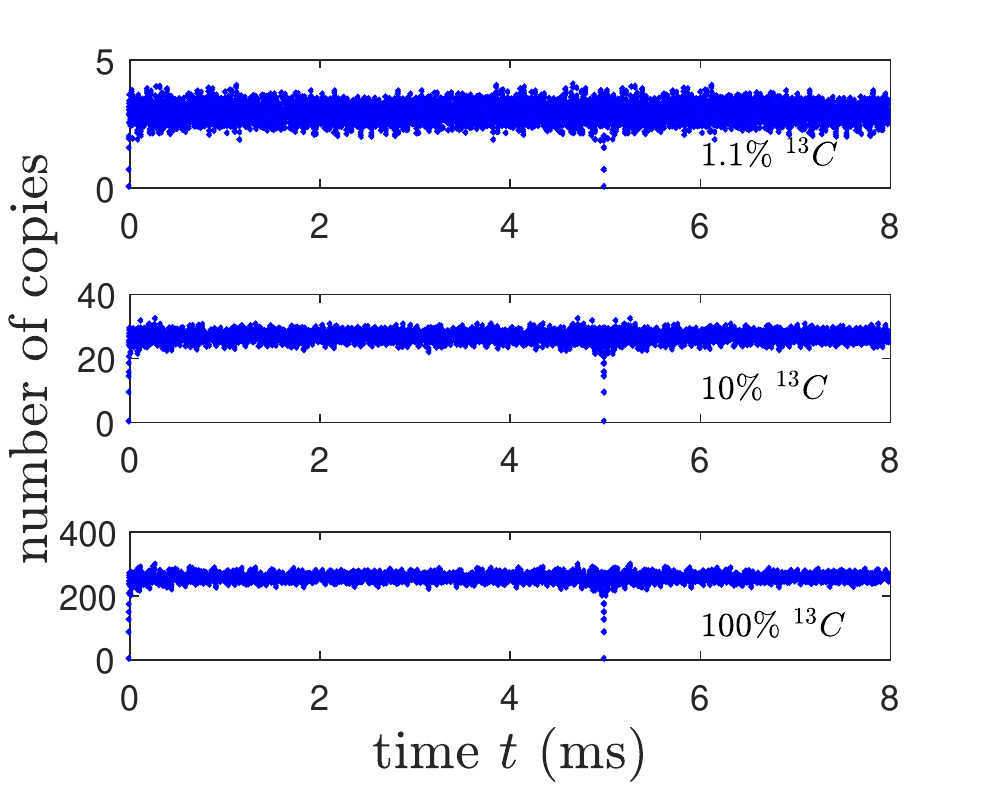, width=0.5\linewidth}}
\caption{Number of copies of information versus the interaction time for various $^{13}C$ concentrations.}
\label{fig:S6}
\end{figure}

Based on recently published hyperfine data of a 510 nuclear spin cluster~\cite{Nizovtsev18}, we show in Fig.~\ref{fig:S6} the number of central spin records over interaction time. The amount of information is estimated using quantum Chernoff information~\cite{Zwolak14,zwolak16}. The results are averaged over ten different realizations of the specified settings. In a highly concentrated $^{13}C$-diamond, a large amount (hundreds) of central spin records could be observed within a timescale of $\mu s$ when the initial nuclear spin polarization is large. At  high $^{13}C$ concentration unwanted  interaction (here in our analysis neglected) between nuclear spins will appear, but our current results show, that on the timescale ($\approx \mu s$) where redundancy increases, these interactions ($\approx$ Hz - kHz) can be neglected for a moderate $^{13}C$ enrichment and a high initial nuclear spin polarization.

\bibliographystyle{apsrev}
\bibliography{reference}